\documentclass[aps,pra,twocolumn,superscriptaddress, biblatex]{revtex4-2}

\pdfoutput=1
\usepackage[english]{babel}

\usepackage[colorlinks=true, breaklinks=true]{hyperref}
\usepackage{epsfig,graphicx}
\usepackage{indentfirst}
\usepackage{amsfonts,amssymb,latexsym,amsmath,enumerate,amsthm}
\usepackage{bm}
\usepackage{color}
\usepackage{appendix}

\usepackage{dcolumn}
\usepackage{centernot}

\usepackage{diagbox}

\usepackage{courier}

\usepackage{url}

\usepackage{physics}
\usepackage{mathtools}

\usepackage{soul}

\usepackage{siunitx}

\usepackage[sort&compress]{natbib}
\begin{document}

\title{Sturmian basis set for the Dirac equation with finite nuclear size: \\ Application to polarizability, Zeeman and hyperfine splitting, \\ and vacuum polarization}
% \title{Sturmian basis for finite size nucleus calculations. \\ Application to vacuum polarization and hyperfine splitting}
% \title{Sturmian basis for relativistic finite size nucleus calculations. \\ Application to vacuum polarization and hyperfine splitting}

\author{V.\,K.~Ivanov}
    \email[E-mail: ]{vladislav.ivanov@metalab.ifmo.ru}
\affiliation{School of Physics and Engineering, ITMO University, 
197101 St.\,Petersburg, Russia}
% \affiliation{School of Physics and Engineering, Faculty of Physics, ITMO University, Kronverksky Pr. 49, bldg. A, St. Petersburg, 197101, Russia}

\author{ D.\,A.~Glazov}
\affiliation{School of Physics and Engineering, ITMO University, 
197101 St.\,Petersburg, Russia}

\author{ A.\,V.~Volotka}
\affiliation{School of Physics and Engineering, ITMO University, 
197101 St.\,Petersburg, Russia}

\date{\today}

\begin{abstract}
We investigate the application of the Sturmian basis set in relativistic atomic structure calculations. We propose a simple implementation of this approach and demonstrate its ability to provide various quantities for hydrogen-like ions, including binding energies, static dipole polarizability, $g$ factor, hyperfine splitting, and nuclear magnetic shielding. Finally, we calculate the all-order (Wichmann-Kroll) vacuum polarization charge density, which was a challenge for the finite-basis-set approach until recently. Comparison of the obtained results with the previously published numerical and analytical calculations is presented. All calculations are performed with the finite size of the nucleus and can in principle be extended to arbitrary binding potentials.
\end{abstract}

\maketitle

\section{Introduction}

Finite basis sets play an important role in atomic and molecular calculations. This versatile approach is used, for instance, in Rayleigh-Ritz method (also known as Ritz-Galerkin method) to form a trial function in the variational procedure. The choice of a particular basis depends on the problem under consideration. In particular, the B-splines with different kinetic balance conditions such as dual-kinetic balance (DKB) are widely used to solve the Dirac equation~\cite{Johnson1986, Sapirstein1996, Shabaev_DKB, Grant2009}. Thus, this basis gained broad application in the field of relativistic quantum mechanics and quantum electrodynamics (QED)~\cite{Volotka2013, Shabaev2024}.
QED effects are important for accurate description of atomic spectra, especially, in highly charged ions. To date, accurate calculations of the one- and two-loop QED effects have been developed for the binding energies (Lamb shift)~\cite{Yerokhin2015, Yerokhin2019}, Zeeman splitting~\cite{Glazov2010, Shabaev2015, Yerokhin2020, Glazov2023, Agababaev2025}, hyperfine splitting~\cite{Yerokhin2005}, transition probabilities~\cite{Volotka2006, Kozlov2024}, and other properties.
% , such as electron self-energy, vacuum polarization (VP) and one-photon exchange \cite{Mohr1998},
% Such effects, in the lowest order, are electron self-energy, vacuum polarization (VP) and one-photon exchange \cite{Mohr1998}. The assumption of QED effects is important to describe the Lamb shift in hydrogen-like ions (see \cite{Yerokhin2019}) and many other effects in atomic spectra.
In quantum chemical calculations, the Gaussian basis is usually more convenient, while it is also widely used in atomic physics, see e.g. a recent development in self-energy calculation in Ref. \cite{Ferenc2025}. Other possible choices of basis sets have been considered, see, e.g., \cite{Quiney2006, grant2007relativistic}.

An interesting alternative to the above-mentioned basis sets is the Coulomb Sturmian basis (or simply Sturmians) \cite{Rotenberg1962, Rotenberg1970, Truskova1984, Sherstyuk1983, Sherstyuk1999, Krylovetsky2001} and its relativistic analogs, such as $L$-spinors of Grant and Quiney \cite{Grant2000} or the Dirac-Coulomb Sturmians of Szmytkowski \cite{Szmytkowski1997}. This basis  yields simple analytic expressions for matrix elements, similarly to the exponential-type basis sets. At the same time, it does not suffer from the linear dependency problem, which is very important in practical calculations. Coulomb Sturmians and their relativistic analogs are constructed to solve equations with point-like nuclear potential. However, the assumption of a finite nuclear size is essential in many cases, such as hyperfine splitting and vacuum polarization. Conversely, calculation of the finite size effects can help to determine the nucleus size (see e.g. \cite{Sun2025}).

The equation for the Sturmian functions allows a potential function of a general form \cite{Rotenberg1962}, so generally it is possible to obtain Sturmian functions for the finite-size nucleus. However, as we shall discuss in this work, such an approach can be complicated.
We propose an alternative approach by relaxing the condition for the basis obeying the Sturmian equation. We impose the relativistic basis, which satisfies the correct asymptotic conditions at zero. In the constructed basis, the common parameter $\lambda$ is freely variable, depending on the scale of the problem --- for example, for the vacuum polarization smaller distances are important, and therefore $\lambda$ is chosen large. This is in contrast to the usual approach where this parameter is bound to the ground state energy \cite{Grant2000}. A similar perspective is taken for example in Ref.~\cite{Herbst2019} in quantum chemical calculations, where the common parameter $\lambda$ is also considered free.

The motivation for this study was raised by recent development in calculations of the vacuum polarization (VP) using finite basis set methods. While B-splines with kinetic balance conditions have been very successful in applications to the diagrams including the electron self-energy and the photon exchange, they failed to provide accurate results for the VP loop beyond the Uehling approximation. On the other hand, in the recent paper by Salman and Saue \cite{Salman2023} and later by Ivanov and coauthors \cite{Ivanov2024}) the Gaussian basis set was successfully used for the calculation of the all-order $(Z\alpha)^{3+}$ (Wichmann-Kroll) VP contribution. A major caveat in the Gaussian basis set is its linear dependence, growing with the number of basis functions. While it can be overcome by enhanced precision algorithms, it makes further progress of this approach problematic. Meanwhile, the Sturmian functions are virtually free of this problem, while enjoying the similar asymptotics at zero and infinity and providing simple algebraic expressions for the matrix elements.

In this work, we develop the relativistic finite basis set with Sturmian functions and demonstrate its abilities in practical calculations. First, we present the necessary theory of the Coulomb Sturmians (CS). Then, we analyze the Dirac wave functions asymptotics at zero and propose a modification of the CS functions for the relativistic case. With this basis set constructed, various contributions for hydrogen-like ion are calculated to test this approach. First, the electron binding energies with the finite-size nuclear model are presented and compared with the results obtained within the DKB method~\cite{Shabaev_DKB}. Next, we test the completeness of the basis set by calculating the static dipole polarizability and compare it with Refs.~\cite{Szmytkowski1997, Grant2000}. Then the first-order hyperfine splitting (HFS), the $g$ factor and the nuclear magnetic shielding (HFS correction to the $g$ factor) are considered to test the approach with matrix elements of different behavior. Finally, we calculate the VP charge density and analyze the results for both close and far distances from the nucleus.

In this paper, we shall use the relativistic unit system $\hbar = c = m = 1$ and the Heaviside charge unit [$\alpha = e^2/(4\pi)$, $e<0$]. 

\section{Basis set}

\subsection{Coulomb Sturmians}

Let us review the theory describing the Coulomb Sturmian (CS) functions. We follow the notation of Grant \cite{Grant2000} below. The Coulomb Sturmians are defined as the solution to a Sturm-Liouville problem:
\begin{equation}
    \left[-\frac{\dd^2}{\dd r^2} + \frac{l(l+1)}{r^2} - 2E_0 + \alpha_{n,l}V(r) \right]S_{nl}(r) = 0
    \label{sturm_eq}
\end{equation}
($E_0$ and $\alpha_{n,l}$ are parameters) with boundary conditions
\begin{equation}
    \left.S_{nl}(r)\right|_{r=0} = \left.S_{nl}(r)\right|_{r\rightarrow \infty} = 0.
\end{equation}
The solution to this problem for the Coulomb potential $V(r) = - \frac{Z\alpha}{r}$ is
\begin{equation}
    S_{nl}(x) = \mathcal{N}_{nl}s_{nl}(x),
    \label{sturm_fun}
\end{equation}
\begin{equation}
    s_{nl}(x) = x^{l+1}e^{-x/2} L_{n-l-1}^{2l+1}(x),
\end{equation}
\begin{equation}
    \mathcal{N}_{nl} = \sqrt{\frac{(n-l-1)!}{2n(n+l)!}},
\end{equation}
where $x=2\lambda r$, $\lambda = \sqrt{-2E_0}$, $L_n^k(x)$ is the Laguerre polynomial. These functions resemble the usual solution of the Schr\"odinger equation; in fact, 
they coincide for the ground state,
if we choose $E_0$ being equal to the ground state energy. The contrast to the Schr\"odinger equation is that for different functions in the spectrum we vary $\alpha_{n,l}$ instead of the energy $E_n$, which is fixed in \eqref{sturm_eq}. We use the following definition of the generalized (associated) Laguerre polynomial:
% (defined for integer parameters, for noninteger $k$ one may use definition via hypergeometric function)
%
\begin{equation}
    L_n^k (x) = (-1)^k \frac{\dd^k}{\dd x^k} L_{n-k} (x), 
\end{equation}
\begin{equation}
    L_{n} (x) = \sum_{m=0}^n \frac{(-1)^m}{m!} \binom{n}{m} x^m.
\end{equation}
These functions are zero at $r=0$ and $r=\infty$, ensuring that the boundary conditions are satisfied. 

The CS functions can be normalized in several ways. Often they are normalized with respect to the $1/r$ weight function (as in \cite{Rotenberg1970}), but for the application of the Rayleigh-Ritz method, it is convenient to normalize them without this weight, meaning that the overlap (Gram) matrix has its main diagonal elements equal to one. Namely,
\begin{equation}
    \int\limits_0^\infty S_{nl}^2(x) \dd x  = \braket{nl}{nl} \equiv g_{nn}^l = 1,
\end{equation}
\begin{equation}
    g_{n,n+1}^l = g_{n+1,n}^l = -\frac{1}{2} \sqrt{1-\frac{l(l+1)}{n(n+1)}}.
\end{equation}
Other elements vanish, so the Gram matrix is tridiagonal  (see \cite{Grant2000} for the derivation via the generating function). 
% Note, that when integrating in respect to $r$ one should divide it by $2\lambda$.
CS functions are orthogonal with respect to $1/x$, so we have the following useful expression:
\begin{equation}
    \bra{nl} \frac{1}{x} \ket{n'l} = \mathcal{N}_{nl}^2\frac{(n+l)!}{(n-l-1)!} \delta_{nn'} = \frac{1}{2n}\delta_{nn'}.
    \label{sturm_orthog}
\end{equation}

\subsection{Radial Dirac equation}

We consider a one-particle relativistic system with a static external electric field $V(\vb{x})$, which is described by the stationary Dirac equation
\begin{equation}
    h_D \phi_n (\vb{x}) = E_n \phi_n(\vb{x})
\label{dirac_eq_2}
\end{equation}
with Hamiltonian
\begin{align}
    h_D =& -i\boldsymbol{\alpha}\nabla + \beta + V(\vb{x}), \label{wave_1}\\    
    \beta =& \gamma^0, \\
    \boldsymbol{\alpha} =& \gamma^0 \boldsymbol{\gamma} = \begin{pmatrix}
        0 & \boldsymbol{\sigma} \\
        \boldsymbol{\sigma} & 0
    \end{pmatrix},
\label{wave_2}
\end{align}
% \begin{align}
%     h_D =& -i\vb{\alpha}\nabla + \beta + V(\vb{x}), \label{wave_1}\\    
%     \beta =& \gamma^0,\, \boldsymbol{\alpha} = \gamma^0 \boldsymbol{\gamma}.
% \label{wave_2}
% \end{align}
%
where $\gamma$ and $\boldsymbol{\sigma}$ are Dirac and Pauli matrices, respectively. We are interested in solutions for systems with spherical symmetry. In this case, the wave function can be expressed in the following way:
\begin{equation}
    \phi_n (\vb{x}) = \frac{1}{r}
    \begin{bmatrix}
        P_{n,\kappa}(r) \Omega _{\kappa,m_j}(\theta,\varphi) \\
        i Q_{n,\kappa}(r) \Omega _{-\kappa,m_j}(\theta,\varphi)
    \end{bmatrix},
\label{anzatz}
\end{equation}
where $P$ and $Q$ are the large and small components of the electron wave function, $\Omega_{\pm \kappa m_j}$ is the spherical spinor (see e.g. \cite{Drake2006hba}), $\kappa = \pm 1, \pm 2, ...$ is the relativistic angular quantum number, $m_j$ is the projection of the total angular momentum. The relation between the orbital angular momentum $l$ and $\kappa$ can be expressed as
\begin{equation}
    l = \abs{\kappa + \frac{1}{2}} - \frac{1}{2}.
\end{equation}

By substituting the \eqref{anzatz} to the \eqref{dirac_eq_2}, we would find that the angular part is factorized, leading to the equation for the radial term of the wave function: 
\begin{equation}
    \begin{bmatrix}
        1 + V & -\frac{d}{dr} + \frac{\kappa}{r} \\
        \frac{d}{dr} + \frac{\kappa}{r} & - 1 + V
    \end{bmatrix}
    \begin{bmatrix}
        P_{n,\kappa} \\
        Q_{n,\kappa}
    \end{bmatrix}
    =
    E_n \begin{bmatrix}
        P_{n,\kappa} \\
        Q_{n,\kappa}
    \end{bmatrix}.
    \label{dirac_eq_mat}
\end{equation}
%
% Later we shall use this matrix equation to obtain the quantities of interest. The solution will be found using variational methods discussed later in the paper.

\subsection{Asymptotics}

% CS
The Coulomb Sturmian basis functions are defined for the Coulomb potential. This approximation is perfect for many applications, but in some problems one wants to assume the finite size of the nucleus. In such a case, the nucleus potential does not contain a singularity, and the asymptotics of the wave function will be different from the Coulomb case.
% The case of the finite nucleus potential is commonly studied along with the use of the Gaussian and Slater basis sets (see e.g. \cite{grant2007relativistic}).
% In the interior of the finite nucleus, we have $\gamma \rightarrow \abs{\kappa}$, so the power of $r$ becomes an integer.
We can show \cite[Sec. 5.4]{grant2007relativistic}, that
\begin{equation}
    \begin{aligned}
        &P \sim r^{l+1} , \, Q \sim r^l, \, &\kappa > 0 \\
        &P \sim r^{l+1} , \, Q \sim r^{l+2}, \, &\kappa < 0
    \end{aligned}
\end{equation}
These two relations can be compactly rewritten for any sign of $\kappa$:
\begin{equation}
    P \sim r^{l_L+1} , \, Q \sim r^{l_S+1},
    \label{pq_asym}
\end{equation}
\begin{equation}
    l_{L,S} = \abs{\kappa \pm \frac{1}{2}} - \frac{1}{2}.
    % l_{L,S} = \sqrt{\abs{\kappa + \frac{1}{2}} + \frac{1}{2}}
\end{equation}

% To perform this with Sturmians, we have to change the power of $r$. Then we have an alternative whether to use standard (non-relativistic) Sturmians or use some of its relativistic versions (for instance, L-spinors). However, 
% ...

Since we want to solve the Dirac equation, we should choose basis functions for the evaluation of $P$ and $Q$.
% The known relativistic Sturmian basis sets are defined for the Coulomb potential. For a finite nucleus, however, the wave function has different asymptotics.
Let us construct a relativistic basis with correct asymptotics at zero. We notice, that \eqref{sturm_fun} has the following asymptotics at zero:
\begin{equation}
    S_{nl}(r) \sim r^{l+1}.
\end{equation}
Comparing the above with \eqref{pq_asym}, we define
\begin{equation}
    \pi_L = S_{n,l_L}(r) , \, \pi_S = S_{n,l_S}(r) .
    \label{cs_basis}
\end{equation}
This approach is similar to one is used for exponential-type basis sets, where transition to the finite nucleus case can be done by a substitution $\pi_{L,S} = r^{\gamma}e^{-b_i r^p} \rightarrow r^{l_{L,S}+1}e^{-b_i r^p}$.
% \footnote{The exponential-type basis functions are defined as $\pi_{L,S} = r^{l_{L,S}+1}e^{-b_i r^p}$, $p=1,2$}.

The only parameters one can vary in the CS basis are its size and $\lambda$. 
Usually, one chooses
% For the Coulomb potential calculations, one usually chooses
$\lambda = \sqrt{-2E_0}$ with $E_0$ being equal to the ground energy (see e.g. \cite{Grant2000}), so that the first function in the basis coincides with the ground state wave function. However, since we are considering the finite nucleus model, the exact solutions cannot be described by a single CS function and $\lambda$ becomes a free parameter, that can be varied depending on the problem.
% A similar approach is used for the point-like nucleus, since, obviously, for heavy nuclei ...
We cover this topic in the Section \ref{section_results} below.

At the end of the section, let us a possible alternative, namely to construct the Strumian basis by substituting the finite-nucleus potential into  Eq. \eqref{sturm_eq}. Since the equation for Sturmians resemble the Schr\"odinger's one (or Dirac, see \cite{Szmytkowski1997, Grant2000}), we use here the same argument that applies to solving this equation for the finite-nucleus case. To solve the equation, we would have to solve it in two regions -- inside and outside the nucleus, and then to match the solutions and their first derivative. Keeping only the regular solution, we would have the wave function expressed via Whittaker W-functions outside the nucleus (see e.g. \cite{Mohr1998} for relativistic solutions), with noninteger parameters (see e.g. \cite{Hannesson1979}). 
\begin{equation}
    \nu = n_r + \delta,
\end{equation}
where $n_r$ is the relativistic quantum number and $\delta$ is a real parameter that vanishes when we set the nucleus size to zero. Finding $\nu$ involves solving a transcendental equation. To employ Sturmians, found this way, one has to either sacrifice the continuity of the first derivative and use Laguerre polynomials with integer parameters outside the nucleus or to tabulate the parameters $\nu$ and deal with W-functions. This approach is possible, while we decide not to take it in this paper due to its complexity.

\section{Rayleigh-Ritz method}

% \subsection{Variation method}

Let us approximate the energy eigenvalues with the next expression:
\begin{equation}
    \varepsilon = \frac{\bra{\Psi}H\ket{\Psi}}{\braket{\Psi}{\Psi}} = \frac{\sum v_i^* v_j H_{ij}}{\sum v_i^* v_j C_{ij}},
\end{equation}
where
\begin{equation}
    \Psi(r) = \sum\limits_{i=1}^n v_i \pi_i(r)
    \label{test_fun}
\end{equation}
are some test functions. These functions can be seen as functions of many parameters
% : $\Psi(v_1,...,v_n,r)$
, by varying which one can approximate the spectrum.
% -- the common method in quantum mechanics
The above definition allows the variation principle to be written in the next form:
\begin{align}
    \pdv{\varepsilon}{v_k^*} =& \frac{\sum v_j H_{kj}}{\sum v_i^* v_j C_{ij}} - \frac{\sum v_i^* v_j H_{ij}\sum v_j C_{kj}}{\left(\sum v_i^* v_j C_{ij}\right)^2} \nonumber \\
    =& \frac{\sum v_j \left(H_{kj} - \varepsilon C_{kj}\right)}{\sum v_i^* v_j C_{ij}}, 
    \label{var_eq}
\end{align}
leading to secular equations
\begin{equation}
    v_j \left(H_{kj} - \varepsilon C_{kj}\right) = 0.
\end{equation}
These can be conveniently written in the matrix form:
\begin{equation}
    \vb{H} \vb{v} = \varepsilon \vb{C} \vb{v}
    \label{eig_eq}.
\end{equation}
For the Dirac equation, the matrices $\vb{H}$ and $\vb{C}$ of size $2n \cross 2n$ have a block structure and are symmetric. The wave function can be expressed as
\begin{equation}
    \psi_{n\kappa}(r) = \begin{bmatrix}
        P_{n\kappa}(r) \\
        Q_{n\kappa}(r)
    \end{bmatrix},
    \label{psi_approx}
\end{equation}
\begin{align}
    P_{n\kappa}(r) &= \sum\limits_{i=1}^n p_{n\kappa,i} \pi^+_i(r), \\
    Q_{n\kappa}(r) &= \sum\limits_{i=1}^n q_{n\kappa,i} \pi^-_i(r).
\end{align}
The coefficients $p$ and $q$ can be conveniently combined in the vector $v_i = (p_1,p_2, ... p_n, q_1, q_2, ... , q_n)$.

Besides the spectrum calculation, the found wave functions are applicable as the intermediate states. The approximate form of the Green's function is \cite{Drake1981}
\begin{equation}
    \frac{1}{h_D - z} \simeq \sum_{n} \frac{\ket{\psi_n}\!\bra{\psi_n}}{E_n - z},
\end{equation}
where $\ket{\psi_n}$ are approximated by \eqref{psi_approx}. We shall use such a Green's function later in the Section \ref{section_results}.

% \subsection{The matrices}

% Let us take a closer look closer at Eq.\eqref{eig_eq}. 
Calculating the matrix elements of the radial Dirac equation \eqref{dirac_eq_mat} with respect to the basis functions, we find for Eq.\eqref{eig_eq}: 
\begin{equation}
    \begin{bmatrix}
        \vb{S}^{LL} + \vb{V}^{LL} & \vb{\Pi}^{LS} \\
        \vb{\Pi}^{SL} & - \vb{S}^{SS} + \vb{V}^{SS} 
    \end{bmatrix}
    \begin{bmatrix}
        \vb{p}_j \\
        \vb{q}_j 
    \end{bmatrix} =
    E_j\begin{bmatrix}
        \vb{S}^{LL} & 0 \\
        0 & \vb{S}^{SS} 
    \end{bmatrix}
    \begin{bmatrix}
        \vb{p}_j \\
        \vb{q}_j 
    \end{bmatrix},
\end{equation}
where
\begin{equation}
    % \vb{S}^{\tau\tau}
    S^{\tau\tau}_{nn'}
    = \int\limits_0^\infty S_{nl_\tau}(2\lambda r)S_{n'l_\tau}(2\lambda r) \dd r = 2\lambda \, g_{nn'}^{l_\tau} ,
\end{equation}
%
% is simply the Gram matrix,
%
\begin{equation}
    % \vb{\Pi}^{LS}
    \Pi^{LS}_{nn'}
    = \int\limits_0^\infty S_{nl_L}(2\lambda r) \left[\frac{\kappa}{r} - \frac{\dd}{\dd r} \right] S_{n'l_S}(2\lambda r) \dd r,
    \label{kinetic_mtx}
\end{equation}
\begin{equation}
    \vb{\Pi}^{SL} = \left( \vb{\Pi}^{LS} \right)^T,
\end{equation}
\begin{equation}
    % \vb{V}^{\tau\tau}
    V^{\tau\tau}_{nn'}
    = \int\limits_0^\infty S_{nl_\tau}(2\lambda r) V(r) S_{n'l_\tau}(2\lambda r) \dd r,
    \label{v_matrix}
\end{equation}
where we symbolically denote $\tau = L,S$.
It is useful to note, that $\vb{\Pi}^{LS}$ is bidiagonal, with its main diagonal elements non-zero for any sign of $\kappa$ and similarly $i=j-2$ elements for $\kappa > 0$ and $i=j+2$ for $\kappa < 0$.

To calculate some matrix elements, including the kinetic matrix $\vb{\Pi}^{LS}$ \eqref{kinetic_mtx}, the following formulas are useful \cite{prudnikov1986integrals}:
\begin{multline}
    \int\limits_0^\infty \dd x \, x^{\alpha-1} e^{-x} L_n^k (x) L_{n'}^{k'}(x) \\= \frac{\Gamma(\alpha) (k' - \alpha -1)_{n'} (k-1)_{n}}{n!n'!} \\
    \cross {}_3F_2 (-n,\alpha,\alpha - k'; k+1, \alpha - k' -n';1) ,
    \label{prud_int}
\end{multline}
% \begin{multline}
%     \int\limits_0^\infty \dd x \, x^{\alpha-1} e^{-cx} L_n^k (cx) L_{n'}^{k'}(cx) \\= \frac{\Gamma(\alpha) (k' - \alpha -1)_{n'} (k-1)_{n}}{c^{\alpha}\, n!n'!} \\
%     \cross {}_3F_2 (-n,\alpha,\alpha - k'; k+1, \alpha - k' -n';1) \\\equiv \frac{1}{c^\alpha} I_{pr} (\alpha; n,k,n',k')
% \end{multline}
%
where $(a)_n$ is a Pochhammer symbol and ${}_3F_2$ is a hypergeometric function. The equivalent formula is \cite{kemble1958fundamental}:
%
% \begin{widetext}
\begin{multline}
    \int\limits_0^\infty \dd x \, x^{p} e^{-x} L_n^k (x) L_{n'}^{k'}(x) \\= (-1)^{n+n'} \,p! \sum_{t=0}^{\min(n,n')} \binom{p-k}{n-t} \binom{p-k'}{n'-t}\binom{p+t}{t} .
\end{multline}
The choice between these two formulas depends on the computation algorithm. In \eqref{prud_int} one should be careful with the cancelling of singularities.

The matrix $\vb{V}^{\tau\tau}$ has simple form when the Coulomb potential $-Z\alpha /r$ is considered: using Eq.\eqref{sturm_orthog} we obtain
\begin{equation}
    \vb{V}^{\tau\tau}_\mathrm{Coul} = -\frac{Z\alpha}{2n}\delta_{nn'}.
\end{equation}

\section{Results}\label{section_results}

In this section, we provide several tests for the proposed basis set. 
We provide calculations with B-spline basis with dual-kinetic balance (DKB) \cite{Shabaev_DKB} (referred below as DKB BS) and with our proposed Coulomb Sturmian basis \eqref{cs_basis} (referred as CS). The computations with Sturmian basis were performed using Python with standard float precision. We used mpmath \cite{mpmath} for some special function evaluation. 
For the B-splines, the Fortran 77 program was used.

For the fine-structure constant, we use the value $\alpha = 1/137.035\,999\,11$ from CODATA 2022 \cite{mohr2024codata}.

We consider the point-like (Coulomb), shell-like and homogeneously charged sphere models of the nucleus. For point nucleus $V_\mathrm{Coul}(r) = -Z\alpha/r$, for shell nucleus
\begin{equation}
    V_\mathrm{shell}(r) = \begin{cases}
        - Z\alpha/r_n, \, &r\leq r_n \\
        - Z\alpha/r, \,\,\, &r > r_n,
    \end{cases}
\end{equation}
where $r_n = \expval{r^2}^{1/2}$ is the mean square size of the nucleus. For sphere model of the nucleus, we have:
\begin{equation}
    V_\mathrm{sphere}(r) = \begin{cases}
        - \frac{Z\alpha}{2R_n}\left(3 - \frac{r^2}{R_n^2}\right) , \, &r\leq R_n \\
        - Z\alpha/r, \,\,\, &r > R_n,
    \end{cases}
\end{equation}
where $R_n = \sqrt{5/3} \expval{r^2}^{1/2}$.
%
% We consider the shell model of the nucleus:
% %
% \begin{equation}
%     V_\mathrm{shell}(r) = \begin{cases}
%         - Z\alpha/r_n, \, r\leq r_n \\
%         - Z\alpha/r, \,\,\, r > r_n,
%     \end{cases}
% \end{equation}
% %
% where $r_n = \expval{r^2}^{1/2}$ is the mean square size of the nucleus. Such a model usually provides good results, since for many effects the $r_n$ is the main parameter of the nuclear charge distribution, that affects the quantities.
The numerical evaluation of Eq.\eqref{v_matrix} can be facilitated by noting, that
% The numerical evaluation of the integrals, including this potential, can be simplified in the following way:
%
\begin{multline}
    V^{\tau\tau}_{nn'} = \int\limits_0^\infty S_{nl_\tau}(2\lambda r) V_\mathrm{shell}(r) S_{n'l_\tau}(2\lambda r) \dd r  
    = -\frac{Z\alpha}{2n}\delta_{nn'} \\
    % = \int\limits_0^\infty S_{nl_\tau}(x) V_\mathrm{Coul}(r) S_{n'l_\tau}(x) \dd r \\
    + \int\limits_0^{r_n} S_{nl_\tau}(2\lambda r) (V_\mathrm{shell}(r) - V_\mathrm{Coul}(r) ) S_{n'l_\tau}(2\lambda r) \dd r.
\end{multline}
% \begin{multline}
%     \vb{V}^{\tau\tau} = \int\limits_0^\infty S_{nl_\tau}(2\lambda r) V_\mathrm{shell}(r) S_{n'l_\tau}(2\lambda r) \dd r  
%     = -\frac{Z\alpha}{2n}\delta_{nn'} \\
%     % = \int\limits_0^\infty S_{nl_\tau}(x) V_\mathrm{Coul}(r) S_{n'l_\tau}(x) \dd r \\
%     + \int\limits_0^{r_n} S_{nl_\tau}(2\lambda r) (V_\mathrm{shell}(r) - V_\mathrm{Coul}(r) ) S_{n'l_\tau}(2\lambda r) \dd r.
% \end{multline}
%
% The first integral is not-zero only for diagonal elements, the second is easier to integrate numerically than one with infinite upper limit.

We note, that no linear dependency problem occurred during the calculations. This is to be expected, since Sturmians are known for this feature (see, for example, the analysis in \cite{Grant2000}). We still controlled this by ensuring the $\vb{B}$-orthogonality (see footnote in \cite{Ivanov2024}), where $\vb{B}$ defined as in $\vb{A}\vb{v}=\lambda\vb{B}\vb{v}$ problem, where matrices $\vb{A}$ and $\vb{B}$ are both symmetric or Hermitian. If $\vb{B}$ (Gram matrix) is a positive-definite matrix, then $\vb{v}_i^T \vb{B} \vb{v}_j = \delta_{ij}$, 
% which is very well satisfied in the present calculations. 
which is fulfilled almost to machine precision in the present calculations.
% which is very satisfied in the present calculations almost up to machine precision. 

\subsection{Binding energy}

First, we present the energies of the electron states in the hydrogen-like ion for the finite nucleus model, which were obtained by the Rayleigh-Ritz method
% of solving the Dirac equation
. The results are presented in Table \ref{tab:energies}, where we present the correction from the finite nucleus size, $E_\mathrm{fns}$.
% They are compared with the analytical results. 
We compare these with the spectrum of the Dirac equation for point-like potential (see, for example, \cite{berestetskii}),
% . The results are presented in Table \ref{tab:energies}. They are compared with the analytical results. The Dirac Coulomb energies are given by (see, for example, \cite{berestetskii})
%
\begin{equation}
    E_{n,\kappa} = \frac{1}{\sqrt{1 + \left(\frac{Z\alpha}{n-\abs{\kappa} + \gamma}\right)^2}},
    \label{dirac_energy}
\end{equation}
where $\gamma = \sqrt{\kappa^2 - (Z\alpha)^2}$.
% where $\lambda_r = \sqrt{\kappa^2 - (Z\alpha)^2}$.
We also solve the Dirac equation within the dual-kinetic-balance method using the basis functions constructed from B-splines \cite{Shabaev_DKB}. The comparison shows that the obtained results are in a good agreement.
% In the first five rows of the Table \ref{tab:energies}, there is a disagreement in the last four digits.

We note, that the spurious states do not appear in our calculations. This means that (at least to some extent) the proposed basis set can be used without kinetic balance schemes, which can be helpful in evaluating the Green's function.

\begin{table}[!htb]
\footnotesize
\caption{\label{tab:energies}%
The electron state energy (in a.u.), calculated with Eq.~\eqref{dirac_energy} (Coulomb potential), with the B-spline basis (with dual kinetic balance), and with Coulomb Sturmian basis \eqref{cs_basis}. The tin hydrogen-like ion is considered, $Z=50$, shell nucleus model is used, $r_n = 4.655$ fm. Basis sizes are $n=80$ for B-spline basis and $n=150$ for Sturmians.
% , $\lambda = 1.46$.
% \textcolor{red}{ALL PRESENTED DIGITS FOR DKB BS ARE STABLE UP TO MACHINE PRECISION}
% $\lambda = 4 Z\alpha$.
}
\begin{ruledtabular}
\begin{tabular}{c|c|S[table-format=1.5, table-number-alignment=center]S[table-format=1.7, table-number-alignment=center]c}
% \colrule
& $E_{n,\kappa}$, point nucleus&  \multicolumn{3}{c}{$E_\mathrm{fns} \cross 10^{6}$} \\
\hline level& Eq.\eqref{dirac_energy} & {DKB BS} 
& {CS} 
& $\lambda$\\

\hline
$1s_{1/2}$& 0.931 059 404 06 & 3.84335
& 3.84335(26)& 2.74\\
$2s_{1/2}$& 0.982 613 709 46  & 0.54109
& 0.54106(15)& 2.74\\
  $2p_{1/2}$& 0.982 613 709 46  & 0.01466
& 0.01468(1)& 2.19\\
  $2p_{3/2}$& 0.983 218 136 26  & 0.00000
& 0.00000
& 1.46\\
$3s_{1/2}$& 0.992 340 868 29& 0.16132
& 0.16161(6)& 1.46\\
  $3p_{1/2}$& 0.992 340 868 29  & 0.00517
& 0.00518(2)& 1.46\\
  $3p_{3/2}$& 0.992 520 428 00& 0.00000
& 0.00000
& 1.46\\
  $3d_{3/2}$& 0.992 520 428 00& 0.00000
& 0.00000
& 1.46\\
  $3d_{5/2}$& 0.992 576 423 81& 0.00000& 0.00000& 1.46\\
\end{tabular}
\end{ruledtabular}
\end{table}

\subsection{Static dipole polarizability}

Let us test the completeness of our basis set. The first basic test is to check if $\sum_n \braket{a}{n}\!\!\braket{n}{b} = \delta_{ij}$, where $\sum_n\ket{n}\!\!\bra{n}$ is constructed with the Rayleigh-Ritz procedure with the CS basis. In matrix form, this can be expressed as follows:
\begin{equation}
    \sum_{ijlk} v_i^{(a)} S_{ij} \left(\sum_n v_j^{(n)} v_l^{(n)} \right)_{jl} S_{lk} v_k^{(b)},
\label{ident_res}
\end{equation}
% \begin{equation}
% \begin{aligned}
%     & (\vb{v}^{(a)})^\mathrm{T} \, \vb{S} \, \vb{I} \, \vb{S} \, \vb{v}^{(b)}  \\
%     & = \sum_{ijlk} v_i^{(a)} S_{ij} \left(\sum_n v_j^{(n)} v_l^{(n)} \right)_{jl} S_{lk} v_k^{(b)},
% \end{aligned}
% \label{ident_res}
% \end{equation}
%
where $\vb{v}^{(n)}$ denotes an eigenvector, corresponding to the $n$-th eigenvalue. Computations show, that the above expression yields the Kronecker delta up to machine precision. It is easy to see that this follows from the B-orthogonality, mentioned earlier in this text.
% \eqref{ident_res} follows from the B-orthogonality, mentioned earlier in this text.

Now we shall use the basis set to evaluate the Green's function. In particular, following Refs.~\cite{Grant2000} and \cite{Szmytkowski1997} we can calculate the static dipole polarizability. 
The correction due to perturbation by a static electric field $\vb{E}$ is \cite{Szmytkowski1997}
% \begin{equation}
%     e \, \vb{r} \cdot  \vb{E},
%     % V_{1} = e \, \vb{r} \cdot  \vb{E},
% \end{equation}
% If we consider the second order perturbation \cite{Szmytkowski1997}
%
\begin{equation}
    \Delta E = - \frac{1}{2} \alpha_d \abs{\vb{E}}^2,
\end{equation}
%
% due to perturbation by a static electric field $\vb{E}$
%
% \begin{equation}
%     V_{1} = e \, \vb{r} \cdot  \vb{E},
% \end{equation}
%
where then $\alpha_d$ is the static dipole polarizability,
\begin{equation}
    \alpha_d = 2 \sum_{n \neq a} \frac{\bra{a}\vb{r}\cdot\vb{e}_z\ket{n}\!\bra{n}\vb{r}\cdot\vb{e}_z\ket{a}}{\varepsilon_a - \varepsilon_n}
\,.
\end{equation}
We denote for simplicity $\ket{n} = \ket{n\kappa m_j}$, which is an eigenstate of corresponding radial and angular operators. First, we have to evaluate the angular integrals. This can be processed easily, keeping in mind the orthogonality of the spherical spinors and the following formulas from \cite{Szmytkowski2007}:
\begin{equation}
\begin{aligned}
    \vb{e}_z \cdot \vb{n} \,  \Omega_{\kappa m} (\vb{n}) = &- \frac{2m}{4\kappa^2-1} \Omega_{-\kappa, m} (\vb{n}) \\&+ \frac{\sqrt{\left(\kappa + \frac{1}{2}\right)^2 - m^2}}{\abs{2\kappa + 1}} \Omega_{\kappa+1, m} (\vb{n}) \\ 
    & + \frac{\sqrt{\left(\kappa - \frac{1}{2}\right)^2 - m^2}}{\abs{2\kappa - 1}} \Omega_{\kappa-1, m} (\vb{n}),
\end{aligned}
\end{equation}
\begin{equation}
\begin{aligned}
    \vb{e}_z \cdot (\vb{n}\cross \boldsymbol{\sigma}) \,  \Omega_{\kappa m} (\vb{n}) &= i\frac{4m\kappa}{4\kappa^2-1} \Omega_{-\kappa, m} (\vb{n}) \\&+ i\frac{\sqrt{\left(\kappa + \frac{1}{2}\right)^2 - m^2}}{\abs{2\kappa + 1}} \Omega_{\kappa+1, m} (\vb{n}) \\ 
    & - i\frac{\sqrt{\left(\kappa - \frac{1}{2}\right)^2 - m^2}}{\abs{2\kappa - 1}} \Omega_{\kappa-1, m} (\vb{n}).
\end{aligned}
\end{equation}
By evaluating the angular integrals
% and assuming the selection rules
, we arrive at \cite{Szmytkowski1997, Grant2000}
\begin{equation}
    \alpha_d = \frac{2}{9} \left(\Delta_{+1} + 2 \Delta_{-2}\right),
\end{equation}
where we have to evaluate the radial integrals in
\begin{equation}
    \Delta_\kappa = \sum_n \frac{\bra{0,-1} 
r \ket{n\kappa}\!\bra{n\kappa} r \ket{0,-1}}{\varepsilon_{0,-1} - \varepsilon_{n\kappa}}.
\label{delta_kappa}
\end{equation}

The numerical results for \eqref{delta_kappa} are presented in Table \ref{tab:polarizability}.
% we present numerical results for $\Delta_\kappa$, obtained with finite basis ...
We compare these with the analytical values obtained from \cite[Eqs. (182)-(184)]{Szmytkowski1997}. The calculations are performed both with the shell nuclear model and with the point-like nucleus, which allows a direct comparison with the results of Refs.~\cite{Grant2000, Szmytkowski1997}.

\begin{table*}[!htb]
\footnotesize
\caption{\label{tab:polarizability}%
% \textcolor{red}{MAYBE REMOVE THIS SUBSECTION ENTIRELY?}
Contributions $\Delta_\kappa$ to the static dipole polarizability. The analytical results \cite{Szmytkowski1997} for a point-like nucleus are presented for comparison. Numerical calculations performed for the Coulomb (point) potential and for the finite size nucleus with basis size $n=100$.
% , 
% $\lambda = 1.46$.
% $\lambda = 4Z\alpha = 1.46$ ... l=4 n=50
}
\begin{ruledtabular}
\begin{tabular}{c|c|cS[table-format=1.8, table-number-alignment=center]S[table-format=1.10, table-number-alignment=center]|cS[table-format=1.8, table-number-alignment=center]S[table-format=1.10, table-number-alignment=center]|c}
% S[table-format=1.7, table-number-alignment=center]}
% \colrule
$Z$  &\textrm{$r_n$ (fm)} 
&  \multicolumn{3}{c|}{$(Z\alpha)^4\Delta_{1}$} & \multicolumn{3}{c|}{$(Z\alpha)^4\Delta_{-2}$} & $\lambda$  \\
\hline   && \cite[Eq.(182)]{Szmytkowski1997} 
&{CS, point nucl.}& {CS, shell nucl.} & \cite[Eq.(183)]{Szmytkowski1997} 
& {CS, point nucl.}& {CS, shell nucl.} & \\

\hline
1  & 0.809 & 6.749 531&6.749 532(1)  &6.749 532(1)   & 6.749 676&6.749 672(1)  &6.749 672(1)  & 0.073 \\
10 & 3.024 & 6.703 128&6.703 126(3)  &6.703 137(2)   & 6.717 556&6.717 553(1)  &6.717 564(0)  & 0.73  \\
20 & 3.476 & 6.563 176&6.563 175(3)  & 6.563 236(1)  & 6.620 296&6.620 293(1)  &6.620 356(0)  & 1.46  \\
50 & 4.655 & 5.611 748&5.611 747(11) & 5.612 887(1)  & 5.942 529&5.942 526(8)  &5.943 695(1)  & 4.38  \\
70 & 5.237 & 4.586 085&4.586 081(38) & 4.590 797(9)  & 5.174 405&5.174 402(33) &5.179 371(18) & 7.15  \\
90 & 5.707 & 3.324 546&3.324 568(102)& 3.342 013(24) & 4.160 097&4.160 121(131)& 4.179 456(61)& 10.51 \\
\end{tabular}
\end{ruledtabular}
\end{table*}

\subsection{Hyperfine splitting}

Now we consider the hyperfine splitting (HFS) in hydrogen-like ion. This is one of the effects, where the finite size of the nucleus is quite important.
% \textcolor{red}{magnetic}
The magnetic dipole hyperfine splitting can be written as \cite{Shabaev1994}
\begin{equation}
    \Delta E = \Delta E_F \{A(Z\alpha)(1-\delta)(1-\varepsilon)+x_\mathrm{rad}\},
\end{equation}
% \begin{equation}
%     \Delta E_\mu = \alpha(Z\alpha)^3 F \{A(Z\alpha)(1-\delta)(1-\varepsilon)+x_\mathrm{rad}\},
% \end{equation}
%
where 
% where for brevity we denote by $F$ a factor depending on the nuclear and the electron properties;
where for brevity we denote by $\Delta E_F$ the splitting without corrections in the brackets;
$A(Z\alpha)$ is the relativistic factor,
\begin{equation}
    A(Z\alpha) = \frac{n^3(2l+1)\kappa(2\kappa(\gamma+n_r) - N_{n_r \kappa})}{N_{n_r \kappa}^4 \gamma (4\gamma^2-1)},
    \label{a(za)}
\end{equation}
where
\begin{equation}
    N_{n_r \kappa} = \sqrt{n_r^2 + 2n_r \gamma + \kappa^2},
\end{equation}
and $\delta$, $\varepsilon$ and $x_\mathrm{rad}$ are nuclear charge distribution, Bohr-Weisskopf and radiative corrections, correspondingly. Below, we consider the nuclear magnetic moment as point-like dipole, but keep finite nuclear charge distribution. 
Calculating the HFS correction involves the evaluation of the following matrix element:
\begin{equation}
    \bra{a} V_{\mathrm{hfs}} \ket{a},
\end{equation}
% \begin{equation}
%     \bra{n\kappa m_j} V_{\mathrm{hfs}} \ket{n\kappa m_j},
% \end{equation}
%
where we have an operator
\begin{equation}
    V_{\mathrm{hfs}} = \frac{[\vb{r} \cross \boldsymbol\alpha]_z}{r^3}.
\end{equation}
% \begin{equation}
%     V_{\mathrm{hfs}} = \frac{[\boldsymbol{\alpha \cross \vb{r}}]_i}{r^3}
% \end{equation}
%
Integrating out the angular coordinates again, we arrive at
\begin{equation}
    2(Z\alpha)^3 A(Z\alpha) = \bra{0,-1} V_{\mathrm{hfs}} \ket{0,-1}.
\end{equation}
In Table \ref{tab:hfs1} we present results for the first-order HFS correction calculations for $1s$ electron in hydrogen-like ion. For finite size nucleus, we consider both shell and homogeneously charged sphere models, since this effect is highly sensitive to the nuclear charge distribution.  We compare our results with analytical formula \eqref{a(za)} from \cite{Shabaev1994}. We denote
\begin{equation}
    A_\mathrm{fs}(Z\alpha) \equiv A(Z\alpha)(1-\delta). 
    \label{a(za)fs}
\end{equation}

\begin{table*}[!htb]
\footnotesize
\caption{\label{tab:hfs1}%
The relativistic factor $A(Z\alpha)$ for the first-order hyperfine correction for the 1s electron in the hydrogen-like ion. The nucleus radii (in fm) and parameters $\delta$ are taken from Shabaev (1994) \cite{Shabaev1994}. Analytical results for $A(Z\alpha)$ are given for comparison. The numerical results for DKB B-splines (with basis size $n=80$) and CS basis ($n=100$) are presented.
}
\begin{ruledtabular}
\begin{tabular}{c|S[table-format=2.5, table-number-alignment=center]S[table-format=2.6, table-number-alignment=center]c|S[table-format=2.5, table-number-alignment=center]S[table-format=2.5, table-number-alignment=center]S[table-format=2.6, table-number-alignment=center]c|cS[table-format=2.7, table-number-alignment=center]c}
% \colrule
%  &  \multicolumn{7}{c}{$aaaa$} \\
% \hline 
& \multicolumn{3}{c|}{shell nucleus} &  \multicolumn{4}{c|}{sphere nucleus} &  \multicolumn{3}{c}{point nucleus} \\
\hline 
$Z$ & {$A_\mathrm{fs}(Z\alpha)$, DKB BS}
& {$A_\mathrm{fs}(Z\alpha)$, CS}  
&$\lambda_\mathrm{fs}$  & {$A_\mathrm{fs}(Z\alpha)$, Eq.\eqref{a(za)fs}} & {$A_\mathrm{fs}(Z\alpha)$, DKB BS}
& {$A_\mathrm{fs}(Z\alpha)$, CS}  
&$\lambda_\mathrm{fs}$  & $A(Z\alpha)$, Eq.\eqref{a(za)} & {$A(Z\alpha)$, CS} 
&{$\lambda$}  
\\
\hline
1 & -1.0000& -1.0000(14)&0.2
& -1.0001 & -1.0001 & -1.0000(14)&0.2& -1.0001& -1.0000(39)& 0.2 
\\
10& -1.0068
& -1.0069(5)&  1.8
& -1.0069 & -1.0069
&-1.0069(4)&1.8& -1.0080& -1.0073(5)& 1.8 
\\
  20& -1.0298& -1.0299(2)&  2.5& -1.0299 & -1.0299&-1.0299(2)&2.5& -1.0329& -1.0318(4)& 3 
\\
 50& -1.2224& -1.2223&  7 & -1.2234 &-1.2230& -1.2230&7& -1.2458& -1.2422(17)& 7 \\
 70& -1.5278& -1.5276&5 & -1.5311 &-1.5297&-1.5296&5& -1.6170& -1.6074(38)&15 \\
 90& -2.1645& -2.1633&6 & -2.1789 &-2.1708& -2.1702&6& -2.6094& -2.5302(196)&20 \\
\end{tabular}
\end{ruledtabular}
\end{table*}

\subsection{Zeeman splitting}

Now we consider the first-order energy shift in magnetic field. For spinless nucleus it is described by the bound-electron $g$ factor. The leading-order contribution is given by the following matrix element,
\begin{equation}
    \Delta E = \frac{\abs{e}}{2} B \bra{a} U \ket{a},
\end{equation}
where $U$ denotes the interaction with the magnetic field,
\begin{equation}
    U = [\vb{r} \cross \boldsymbol\alpha]_z .
\end{equation}
The $g$ factor is defined by the following expression,
\begin{equation}
    \Delta E = \frac{\abs{e}}{2} g B M_j .
\end{equation}
$M_j$ here is the total angular momentum projection on $z$-axis directed along the magnetic field $\vb{B}$.
For a hydrogenic ion with point-like nucleus the $g$ factor can be found analytically,
\begin{equation}
    g = \frac{\kappa}{\kappa^2-1/4} \left( \kappa E_{n,\kappa} - \frac{1}{2} \right).
    \label{g_fac}
\end{equation}
% \begin{equation}
%     g = \frac{\kappa}{j(j+1)} \left( \kappa E_{n,\kappa} - \frac{1}{2} \right)
% \end{equation}
The $g$-factor correction due to the finite size of the nucleus can be found by the following formula \cite{Glazov2002, Karshenboim2005}:
% \cite{Zatorski2012}(fix):
%
\begin{equation}
    \delta g_\mathrm{fs} = \frac{4(2\gamma_1+1)}{3} E_\mathrm{fns},
    \label{g_fac_fs}
\end{equation}
where $\gamma_1 = \sqrt{1 - (Z\alpha)^2}$. The numerical and analytical results for the $g$ factor are shown in Table \ref{tab:g_fac}. The results for CS basis are in a good agreement with ones obtained with B-splines and with Eq.~\eqref{g_fac_fs}. We also calculated this value for the sphere nuclear model, however, the deviation from the shell model is tiny and only noticeable for heavy elements, so we do not present it.

\begin{table*}[!htb]
\footnotesize
\caption{\label{tab:g_fac}%
% HFS+magnetic ... 
The $g$ factor for the 1s electron in hydrogen-like ion. The nuclear radii (in fm) are taken from Ref.~\cite{Moskovkin2004}. For comparison, analytical results $g$ for the point-like nucleus and numerical results $g_\mathrm{fs}$ for the extended nucleus calculated with DKB B-splines (basis size $n=80$) and CS basis ($n=100$) are given.
}
\begin{ruledtabular}
\begin{tabular}{c|cccS[table-format=1.13, table-number-alignment=center]|cS[table-format=1.14, table-number-alignment=center]|S[table-format=1.5, table-number-alignment=center]}
% \colrule
 % &  \multicolumn{6}{c}{aaaa} \\
% \hline
 &  \multicolumn{4}{c|}{{shell nucleus}} & \multicolumn{2}{c|}{{point nucleus}} & \\
\hline
$Z$ & \textrm{$r_n$ (fm)} & $g_\mathrm{fs}$, Eqs.\eqref{g_fac}, \eqref{g_fac_fs}& $g_\mathrm{fs}$, DKB BS 
&  {$g_\mathrm{fs}$, CS}
&$g$, Eq.\eqref{g_fac} & {$g$, CS}
&{$\lambda$}  \\

\hline
1 & 0.880& 1.999 964 499& 1.999 964 499 &  1.999 964 499  &1.999 964 499& 1.999 964 499   &0.0584 \\
10& 2.967& 1.996 445 176& 1.996 445 176 &  1.996 445 17  &1.996 445 171& 1.996 445 16   &0.584  \\
20& 3.495& 1.985 723 318& 1.985 723 318 &  1.985 723 3  &1.985 723 204& 1.985 723 2   &1.17  \\
50& 4.643& 1.908 093 900& 1.908 093 760 &  1.908 093 8  &1.908 079 205& 1.908 079    &2.92   \\
70& 5.228& 1.813 056 712& 1.813 056 048 &  1.813 056   &1.812 921 138& 1.812 921   &4.09   \\
92& 5.834& 1.656 133 964& 1.656 121 518 &  1.656 122  &1.654 846 170& 1.654 86 &5.37   \\
\end{tabular}
\end{ruledtabular}
\end{table*}

Further, we consider the hyperfine-interaction correction to the $g$ factor, which is the second-order effect. This involves calculating the nuclear magnetic shielding constant $\sigma$ \cite{Moskovkin2004, Yerokhin2011, Yerokhin2012, volchkova2021}
% , which involves evaluation of the Green's function
. The leading contribution is written as
\begin{equation}
    \sigma_0 = \alpha \sum_{n \neq a} \frac{\bra{a}U\ket{n}\!\bra{n}V_{\mathrm{hfs}}\ket{a}}{\varepsilon_a - \varepsilon_n}.
\end{equation}
%
% where $U$ denotes the interaction with the magnetic field
% %
% \begin{equation}
%     U = [\vb{r} \cross \boldsymbol\alpha]_z .
% \end{equation}
%
This constant can be parametrized as $\sigma_0 = \alpha\,(Z\alpha) S(Z\alpha)/3$ \cite{Moskovkin2004}. For a point-like nucleus and $1s$ electron in a hydrogenic ion $S(Z\alpha)$ is written as,
\begin{equation}
    S(Z\alpha) = \frac{2}{3} \left[\frac{2+\gamma_1}{3(1+\gamma_1)} + \frac{2}{\gamma_1(2\gamma_1-1)} \left(1-\frac{\gamma_1}{2} + (Z\alpha)^2\right)\right].
    \label{s(za)}
\end{equation}
In Table \ref{tab:hfs_mag} we present results for $S(Z\alpha)$, evaluated for the ground state electron in the hydrogen-like ion, using the finite basis set method and via Eq.\eqref{s(za)}. We consider point, shell and sphere nuclus models here. We also present numerical results for extended nucleus from \cite{Moskovkin2004}. Again, we adjust $\lambda$ for different ions. Note, that nucleus radii presented in Tables \ref{tab:hfs1} and \ref{tab:hfs_mag} are different.

\begin{table*}[!htb]
\footnotesize
\caption{\label{tab:hfs_mag}%
% HFS+magnetic ... 
The $S(Z\alpha)$ factor for the hyperfine-correction to the $g$-factor for 1s electron in the hydrogen-like ion. The nucleus radii (in fm) are the same as in Table \ref{tab:g_fac} and taken from Moskovkin \textit{et al.} (2004) \cite{Moskovkin2004}. For comparison, analytical results for the Coulomb potential $S(Z\alpha)$ and numerical results for the extended nucleus $S_\mathrm{fs}(Z\alpha)$ from \cite{Moskovkin2004} are given. The numerical results for DKB B-splines (with basis size $n=120$) and CS basis ($n=100$) are presented.
}
\begin{ruledtabular}
\begin{tabular}{c|S[table-format=1.7, table-number-alignment=center]S[table-format=1.7, table-number-alignment=center]c|S[table-format=1.5, table-number-alignment=center]S[table-format=1.7, table-number-alignment=center]S[table-format=1.7, table-number-alignment=center]c|S[table-format=1.5, table-number-alignment=center]S[table-format=1.9, table-number-alignment=center]c}
% \colrule
 % &  \multicolumn{6}{c}{aaaa} \\
% \hline
& \multicolumn{3}{c|}{shell nucleus} &  \multicolumn{4}{c|}{sphere nucleus} &  \multicolumn{3}{c}{point nucleus} \\
\hline 
$Z$ & {$S_\mathrm{fs}(Z\alpha)$, DKB BS}
& {$S_\mathrm{fs}(Z\alpha)$, CS}
&  $\lambda_\mathrm{fs}$ & {$S_\mathrm{fs}(Z\alpha)$ \cite{Moskovkin2004}} 
& {$S_\mathrm{fs}(Z\alpha)$, DKB BS}
& {$S_\mathrm{fs}(Z\alpha)$, CS}
&$\lambda_\mathrm{fs}$ & {$S(Z\alpha)$, Eq.\eqref{s(za)}} & {$S(Z\alpha)$, CS}
&$\lambda$ 
\\

\hline
1 & 1.00014 &1.00014(3)&0.15 & 1.00014&1.00014&1.00014(3)&0.15 & 1.00014&1.00014(3)    &0.15 
\\
10 & 1.01444&1.01445(2)&1.25 & 1.01444&1.01444 &1.01445(2)  &1.25 & 1.01446&1.01446(2)    &1.25 
\\
20 & 1.05900&1.05903(3)&2    & 1.05901&1.05901 &1.05903(3)  &2    & 1.05927&1.05920(3)    &2 
\\
50 & 1.43427   &1.43420(8)&6    & 1.43471&1.43459    &1.43455(9)  &6    & 1.44624&1.44432(32)   &6
\\
70 & 2.04858   & 2.04842(18)&8  & 2.05090&2.05034    &2.05027(22) &8    & 2.13349& 2.12284(370) &12
\\
92& 3.56943    & 3.57004(80)&10 & 3.58300& 3.57999   &3.58025(12) &10   & 4.37922& 4.21927(2541)&20
\\
\end{tabular}
\end{ruledtabular}
\end{table*}

\begin{figure}[!htb]
    \centering
    \includegraphics[width=1\linewidth]{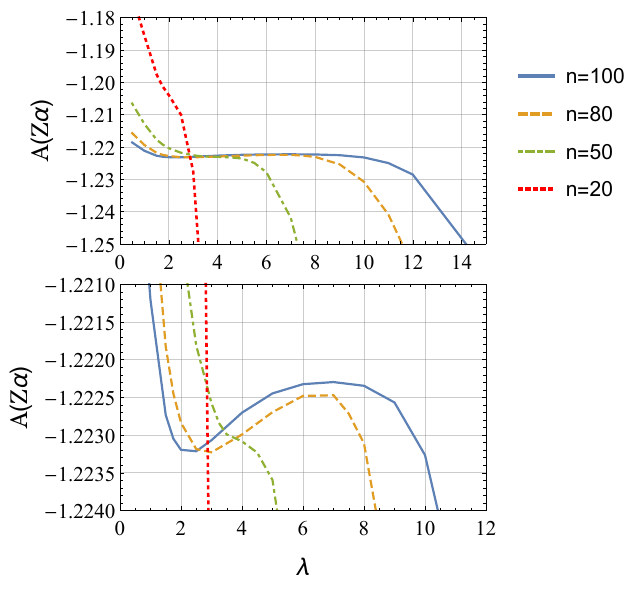}
    \caption{The dependence of the computed value of $A_\mathrm{fs}(Z\alpha)$ on $\lambda$ for $Z=50$ and $r_n = 4.643$ fm and different sizes of the basis.
    }
    \label{fig:stability_hfs1}
\end{figure}

\begin{figure}[!htb]
    \centering
    \includegraphics[width=1\linewidth]{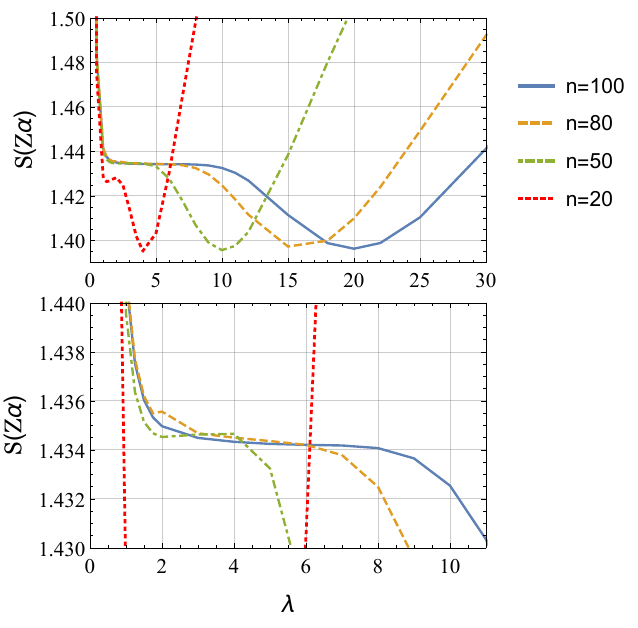}
    \caption{The dependence of the computed value of $S_\mathrm{fs}(Z\alpha)$ on $\lambda$ for $Z=50$ and $r_n = 4.643$ fm and different sizes of the basis.
    }
    \label{fig:stability}
\end{figure}

Let us investigate the stability of the presented results due to the variation of parameter $\lambda$.
We found, that the computed values are stable within a range of $\lambda$ values. We present $A_\mathrm{fs}(Z\alpha)$ and $S_\mathrm{fs}(Z\alpha)$, calculated with different $\lambda$ and $n$ in Figures \ref{fig:stability_hfs1} and \ref{fig:stability} to depict the behavior of the computed values. We can see that the curves have a plateau, which becomes wider with the basis increase (upper part of the figures). Zooming in we see, that the curves in Fig. \ref{fig:stability_hfs1} have a pronounced maximum, by identifying which we can obtain the correct value of $A(Z\alpha)$. However, for lower $Z$ this feature disappears and the $n=100$ curve will look similar to the one corresponding to $n=50$ in Fig. \ref{fig:stability_hfs1} -- then the correct value can be identified by observing the region where the curvature changes rapidly. For $S(Z\alpha)$ we do not observe such minimum, therefore the correct value can be taken as the one corresponding to the plateau. We tabulate the computed values in Table \ref{tab:stability} to illustrate the stability of the computed quantity in the plateau region. The numbers in the Table \ref{tab:stability} are clearly stabilizing around the correct value. The similar picture can be observed for other computed quantities.
% , like the polarizability
The character of the curve's plateau implies the uncertainty of the computed value, which is reflected by numbers in brackets in Tables \ref{tab:polarizability}, \ref{tab:hfs1} and \ref{tab:hfs_mag}.

\begin{table}[!htb]
\footnotesize
\caption{\label{tab:stability}%
The dependence of the computed value of $S_\mathrm{fs}(Z\alpha)$ on the CS basis size $n$ and $\lambda$. The hydrogen-like ion with $Z=50$ and $r_n = 4.643$ fm \cite{Moskovkin2004} is considered.
}
\begin{ruledtabular}
\begin{tabular}{c|cccc}
% \colrule
 % &  \multicolumn{6}{c}{aaaa} \\
% \hline
\diagbox{$\lambda$}{$n$}
& 20& 50& 80& 100\\

\hline
0.1& 2.91001& 3.17697& 3.04282& 2.99222\\
 0.5& 1.47493& 1.48237& 1.48405&1.48348\\
 1& 1.42835& 1.43956& 1.44122&1.44101\\
2& 1.42819& 1.43452& 1.43556& 1.43496\\
 3& 1.41392& 1.43462& 1.43468&1.43449\\
  4& 1.39519& 1.43465& 1.43450& 1.43433\\
 6& 1.43049& 1.42773& 1.43420&1.43420\\
  8& 1.49814& 1.40689& 1.43248& 1.43407\\
10& 1.55188& 1.39561& 1.42472& 1.43253\\
 12& 1.58574& 1.40376& 1.41155&1.42687\\
 15& 1.61007& 1.43837& 1.39720&1.41140\\
  20& 1.61919& 1.50607& 1.40986& 1.39623\\
 30& 1.61656& 1.58502& 1.49210&1.44120\\
 50& 1.61511& 1.60442& 1.58812&1.55627\\
 80& 1.61341& 1.59281& 1.59904&1.59970\\
\end{tabular}
\end{ruledtabular}
\end{table}

% ...

\subsection{Vacuum polarization}
% \subsection{VP Induced charge}

% ...

% We briefly review the formulas for calculating the vacuum polarization for $Z\alpha \geq 3$ order using the finite basis set method; the details can be found in \cite{Salman2023, Ivanov2024}. 
% To demonstrate the applicability of the proposed basis, we shall calculate the induced charge density.
The effect of vacuum polarization is very sensitive to the nucleus model, which means that the difference between the results for the finite nucleus model and the Coulomb model is substantial (see \cite{Rinker1975}).
% -- for the case of a point-like potential, the induced charge density has a delta-function component
The results for both cases, when the finite basis set method is applied, are presented in \cite{Salman2023}. The induced charge density is mostly localized in the vicinity of the nucleus, with its peak reside slightly above the mean-square radius of the nucleus.

The linear in $Z\alpha$ (Uehling) contribution should be renormalized and is usually considered separately. We subtract this term from the VP density -- the remaining part is called the many-potential or Wichmann-Kroll part \cite{Wichmann1956}.

The VP charge density can be expressed as \cite{Mohr1998}
\begin{equation}
\begin{aligned}
    &\rho(\vb{x}) = e \left.\Tr[S_F(x,x')\gamma_0]\right|_{x' \rightarrow x} \\
    &= \frac{e}{2} \left(\sum\limits_{E_n > 0} \phi^\dagger_n(\vb{x}) \phi_n(\vb{x}) \right. - \left. \sum\limits_{E_n < 0} \phi^\dagger_n(\vb{x}) \phi_n(\vb{x})  \right),
    \label{ind_charge}
\end{aligned}
\end{equation}
where $S_F(x,x')$ is the electron propagator in Furry's picture. The limit $x' \rightarrow x$ is assumed to be the mean value of the limits from ``left" and ``right".

The VP density can be expanded into the partial components \cite{Wichmann1956}, which is useful since only the first few components add in a significant part of the VP charge (corresponding wave functions are more localized near the nucleus):
\begin{equation}
    \rho(\vb{x}) = \sum\limits_{\kappa = \pm 1}^{\pm\infty} \rho_\kappa(\vb{x}).
    \label{vp_kappa_dec}
\end{equation}
% \begin{equation}
%     \rho(\vb{x}) = \sum\limits_{\kappa = \pm 1}^{\pm\infty} \rho_\kappa(\vb{x}) = \frac{e}{2\pi i} \int\limits_{C_F} \dd z  \sum\limits_{\kappa \pm 1}^\infty \frac{\abs{\kappa}}{2\pi} \Tr G_\kappa(\vb{x}, \vb{x}, z).
%     \label{vp_kappa_dec}
% \end{equation}

For our problem, the expression for the VP induced charge can be conveniently written as \cite{Salman2023}
\begin{align}
    \rho_\kappa(\vb{x}) =& \frac{\abs{\kappa}}{2\pi}\frac{e}{2}\frac{1}  {r^2}\sum\limits_n \text{sgn}(E_{\kappa,n})\rho_{\kappa,n}(r),
    \label{rho1}
    \\ \rho_{\kappa,n}(r) &= \varphi_{\kappa,n}^\dagger\varphi_{\kappa,n} = P_{n,\kappa}^2 + Q_{n,\kappa}^2.
\end{align}
where index $n$ means infinite sum over energy eigenstates (for the exact propagator) or finite sum for the operator, found by the Rayleigh-Ritz procedure.

We wish to find the regular part of the VP density, therefore we have to, in general, subtract the linear in $Z\alpha$ charge \textit{and} the spurious contribution of order $(Z\alpha)^3$; however, the latter vanishes for individual $\kappa$ terms \cite{Soff1988} and can be neglected as we are interested in the first few $\kappa$ terms. Then
% We are to find that part of VP density, which are not to be renormalized, so in general we have to subtract linear in $Z\alpha$ charge and spurious contribution of order $(Z\alpha)^3$; however, the latter is vanishing for individual $\kappa$ terms \cite{Soff1988}.
%
\begin{equation}
    \rho^{n \geq 3} = \rho - \rho^{(1)}.
\end{equation}
% \begin{equation}
%     \rho^{n \geq 3} = \rho - \rho^{(1)} - \Tilde{\rho}^{(3)}
% \end{equation}

To improve the numerical results we manually enforce the $\mathcal{C}$-symmetry of the VP density \cite{Salman2023} (which is shown to be equivalent to using the dual-kinetic balance for the VP calculation problem)
\begin{equation}
    \rho_{\kappa,\mathcal{C}}(r,Z) \equiv \frac{1}{2}\left(\rho_\kappa(r,Z) - \rho_\kappa(r,-Z)\right).
\end{equation}
Then we have the following expression for the VP charge density \cite{Salman2023, Ivanov2024}:
\begin{equation}
\begin{aligned}
    \rho_{\kappa,\mathcal{C}}^{n \geq 3}(r,Z) &\approx \frac{1}{2}\left(\rho_\kappa(r,Z) - \rho_\kappa(r,-Z)\right) \\
    &- \frac{1}{2}\frac{Z}{\delta}\left(\rho_\kappa(r,\delta) - \rho_\kappa(r,-\delta)\right).
    \label{c_sym}
\end{aligned}    
\end{equation}
where we evaluate a numerical derivative, with the parameter $\delta$ being small (we set $\delta = 10^{-6}$).

% ...

% Finally,
We present the VP charge density for $(Z\alpha)^{\geq 3}$, calculated with the CS basis and Eq.\eqref{c_sym}. 
% The results presented are for the non-linear (many-potential) VP charge density.
We consider a hydrogen-like uranium ion with $Z=92$ and $r_n = 5.8507$ fm, similar to \cite{Ivanov2024}. 
The induced charge is mainly concentrated in the vicinity of the nucleus, 
% therefore we choose the basis set with 
suggesting the following choice of the parameter $\lambda$:
\begin{equation}
    \lambda \sim \frac{1}{r_n}.
\end{equation}

\begin{figure}[!htb]
    \centering
    \includegraphics[width=1\linewidth]{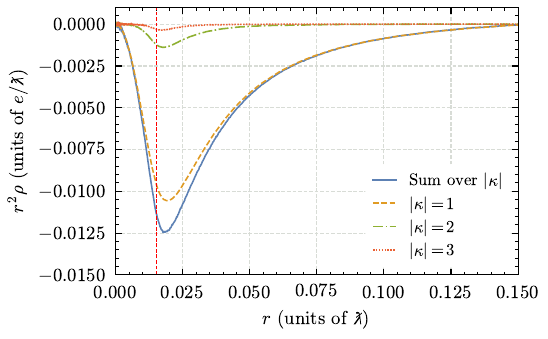}
    \caption{VP induced charge density, calculated with CS basis, $\lambda = 94$. The individual $\abs{\kappa}$ contributions and their sum are presented.
    % vp plot z92 n150 l140
    }
    \label{fig:vp1}
\end{figure}

\begin{figure}[!htb]
    \centering
    \includegraphics[width=1\linewidth]{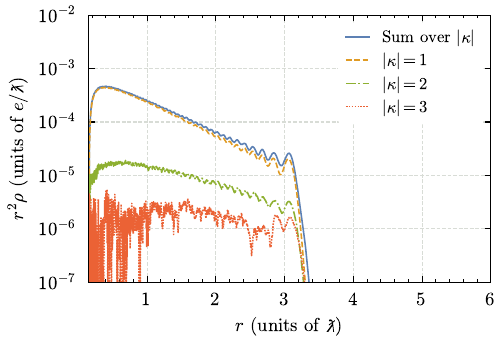}
    \caption{Same as Figure \ref{fig:vp1}, but at large distance, in log-scale.
    % vp logplot z92 n150 l140
    }
    \label{fig:log_vp1}
\end{figure}

\begin{figure}[!htb]
    \centering
    \includegraphics[width=1\linewidth]{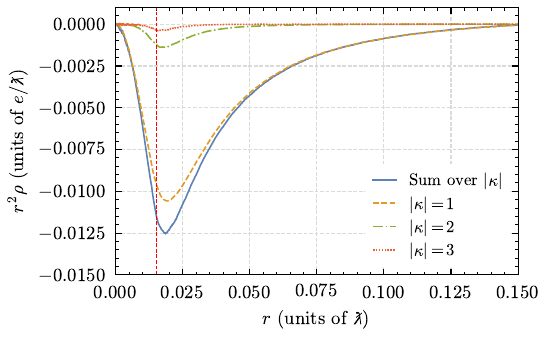}
    \caption{VP induced charge density, calculated with CS basis, $\lambda = 54$.
    % vp plot z92 n150 l80
    }
    \label{fig:vp2}
\end{figure}

\begin{figure}[!htb]
    \centering
    \includegraphics[width=1\linewidth]{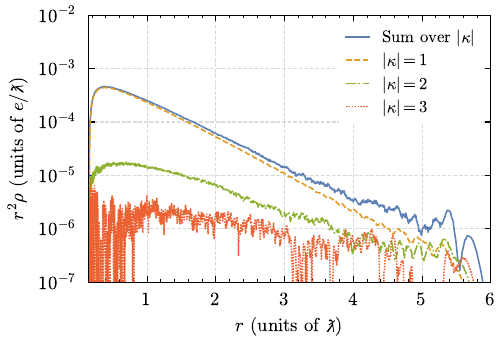}
    \caption{Same as Figure \ref{fig:vp2}, but at large distance, in log-scale.
    % vp logplot z92 n150 l80
    }
    \label{fig:log_vp2}
\end{figure}

\begin{figure}[!htb]
    \centering
    \includegraphics[width=1\linewidth]{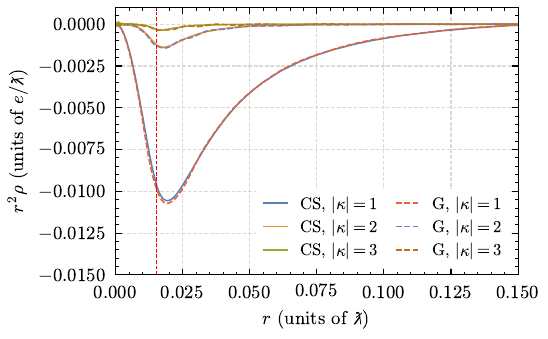}
    \caption{A comparison between VP densities, acquired via CS basis, $\lambda = 94$, and Gaussian basis (denoted "G") with dual-kinetic balance, $n=30$ in Ivanov \textit{et al.} \cite{Ivanov2024}.
    % vp logplot3 z92 n150 l80
    }
    \label{fig:comp1}
\end{figure}

\begin{figure}[!htb]
    \centering
    \includegraphics[width=1\linewidth]{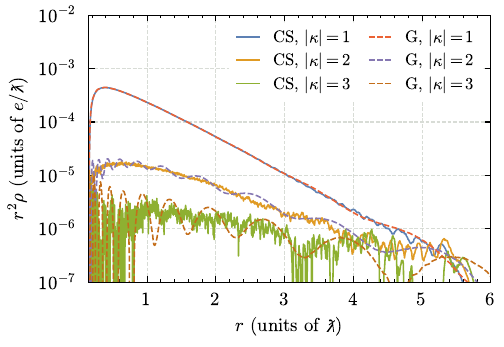}
    \caption{A comparison between VP densities (large distance, log-scale), acquired via CS basis, $\lambda = 54$, and Gaussian basis (denoted "G") with dual-kinetic balance, $n=30$ in Ivanov \textit{et al.} \cite{Ivanov2024}.
    }
    \label{fig:log_comp1}
\end{figure}

We present the results for the close distance from the nucleus $r<0.015 \lambdabar$ ($\lambdabar$ is the Compton wavelength of the electron) in Figs. \ref{fig:vp1}, \ref{fig:vp2} and for the far distance $0.015 \lambdabar < r < 6 \lambdabar$ in Figs. \ref{fig:log_vp1}, \ref{fig:log_vp2}. The obtained results are meaningful for contributions of $\abs{\kappa} \leq 3$. It is interesting to note, that if $\lambda$ is chosen too large, then the curves collapse for $r$ greater than some cut-off distance; the shape of the VP curve near this cut-off resembles the behavior of the Sturmians for large $n$. The results for CS basis are slightly better than ones can be achieved with the Gaussian basis when machine precision is used, see Figs. \ref{fig:comp1}, \ref{fig:log_comp1} and Ref. \cite{Ivanov2024}. For the Gaussian basis, the dual kinetic balance was applied, the basis size was set to $n=30$ (in Ref. \cite{Ivanov2024} it was discussed, that calculations with the Gaussian set collapse if $n$ is chosen too large). The difference is mostly noticeable for the $\abs{\kappa}=2$ contribution. However, it should be noted that the quality of the results reaches a plateau as the basis size $n$ increases, indicating the slowing of the convergence. The curves presented can be compared with those obtained with the finite basis set method of Ref. \cite{Salman2023} and with those calculated with the Green's function integrating in Ref. \cite{Mohr1998}.

\section{Discussion and conclusion}\label{section_v}

We have presented a variant of a Sturmian-like basis set, constructed under the assumption of the correct behavior of the basis functions in the interior of the finite-size nucleus. In the proposed basis, the basis functions for the large and small components of the wave function are constructed from the Coulomb Sturmians by a substituting integers $l_{L,S}$ instead of $l$, providing correct asymptotics at the zero.
We have discussed that the assumption of the finite-nucleus potential in the equation for the Sturmian functions while being possible, is complicated: one would have to tabulate the parameters, that generate the basis, instead of using integers and would be dealing with Whittaker W-functions. This would sacrifice the convenience of the Sturmian-like basis sets, therefore we have taken an alternative route.

We have applied the proposed basis set and the Rayleigh-Ritz method to calculate a variety of quantities in the hydrogen-like ions. First, we obtained an energy spectrum, using this basis set and have compared it with calculations using B-splines with dual-kinetic balance and with analytical expressions (Table \ref{tab:energies}).
Then, we evaluated several corrections, which involve evaluating Green's function and assumed the finite size of the nucleus. First, we calculated the static dipole polarizability for the Coulomb and the shell potential, see Table \ref{tab:polarizability}. These results can be compared with analytical expressions of 
% Grant \cite{Grant2000} and 
Szmytkowski \cite{Szmytkowski1997}. 
We see that the results for the extended nucleus have lesser uncertainty, reflecting the correct choice of the asymptotics inside the nucleus. Since the form of the perturbation operator ($\sim r$), this correction is not largely dependent on the basis function behavior near the zero. The similar picture is observed for the first-order $g$-factor calculation (Table \ref{tab:g_fac}), where CS basis provide good results for both models of the nucleus.

In contrast, the calculations of corrections, which are sensible to the nucleus charge distribution, show the large difference for point and extended nuclei. To probe the behavior of our CS basis at small distances, we have considered hyperfine splitting (HFS) corrections, namely, the first-order HFS correction and the HFS correction to the $g$-factor. These results are presented in Tables \ref{tab:hfs1}, \ref{tab:hfs_mag}, where we considered point-like, shell-like and homogeneously charged sphere models of the nucleus. For light elements ($Z \lesssim 20$) the results for the point and the finite nuclei are of similar accuracy, while for heavier elements there is a noticeable discrepancy between those, showing the sensitivity to the wave function asymptotics at zero. The proposed basis shows good results for the extended nucleus, reflecting the correct choice of the asymptotics. 

Finally, we have shown the results for the vacuum polarization. The B-spline basis set, which is commonly used in atomic calculations, 
is not suitable in this case. Conversely, the results for Coulomb Sturmians are on a par with those for Gaussian basis \cite{Ivanov2024}. The Figures \ref{fig:comp1} and \ref{fig:log_comp1} show, that the CS basis provides slightly better results, especially for the $\abs{\kappa}=2$ contribution. The main strength of the Strumian basis is that it is linearly independent to a high degree (see discussion in Ref. \cite{Grant2000}), which is in sharp contrast to the Gaussian basis, for which the linear dependence imposes an upper limit on the basis size.

\begin{acknowledgments}
We are grateful to V.A. Agababaev, M.G. Kozlov, and A.V. Malyshev for fruitful discussions.
This work was supported by the Ministry of Science and Higher Education of the Russian Federation (Project No. FSER-2025–0012) and by the Foundation for the Advancement of Theoretical Physics and Mathematics ``BASIS''.
\end{acknowledgments}

% \begin{appendix}
% \section{aaa}
% bbb
% \end{appendix}

\bigbreak

\bibliography{main.bib}

\end{document}